\documentclass{article}

\RequirePackage{amsmath}
\RequirePackage{amssymb}
\RequirePackage{latexsym}
\RequirePackage{curves}
\usepackage{siunitx}
\usepackage{amsfonts}
\usepackage{url}
\usepackage{epsfig}

\usepackage[numbers]{natbib}

\begin{document}

\title{Thermal convection in a linearly viscous fluid overlying a bidisperse porous medium}

\author{P. Dondl\\
Abteilung f\"ur Angewandte Mathematik\\ Albert-Ludwigs-Universit\"at Freiburg\\
Hermann-Herder-Str.\ 10,\\
79104 Freiburg, Germany\\
\\
and\\
\\
B. Straughan\\
Department of Mathematics\\ University of Durham, Durham DH1 3LE, UK}

\maketitle

\begin{abstract}
A bidisperse porous medium is one with two porosity scales. There are the usual pores known as macro pores but also cracks or fissures in the
skeleton which give rise to micro pores. In this article we develop and analyse a model for thermal convection where a layer of viscous
incompressible fluid overlies a layer of bidisperse porous medium. Care has to be taken with the boundary conditions at the interface of the fluid
and the porous material and this aspect is investigated. The situation is one in a layer which is heated from below and under appropriate conditions
bimodal neutral curves are found. These depend on the ratio ${\hat d}$ of the depth $d$ of the fluid layer to the depth $d_m$ of the porous layer.
We show that there is a critical value of ${\hat d}$ such that below this value convective motion initiates in the porous layer whereas for ${\hat d}$
above this value the convective instability commences in the fluid layer.
\end{abstract}

\section{Introduction}	
\label{Intro}

The problem of flow of a fluid overlying a porous medium saturated by the same fluid is one which has attracted the attention of many
prominent scientists. A fundamental interface condition between the fluid and the porous medium was proposed by \citet{BeaversJoseph:1967}.
The first analysis of thermal convection in the situation where a fluid overlies a saturated porous medium is due to \citet{Nield:1977}
who succesfully employed the Beavers - Joseph boundary condition to derive a satisfactory model. A surprising result for the same thermal
convection problem was discovered by \citet{ChenChen:1988} who showed that the ratio of fluid depth, $d$, to porous layer depth, $d_m$, 
defined by ${\hat d}=d/d_m$, is critical to determining the process for the onset of thermal convection in the two layer system. If ${\hat d}$
is below a critical value then convection commences in the porous layer whereas when ${\hat d}$ is above the critical value then convection
commences in the fluid. This class of problem was further investigated both experimentally and theoretically by \citet{ChenChen:1989,ChenChen:1992},
\citet{Chen:1991}, \citet{McKayStraughan:1993}, where the last mentioned article applies the theory to the problem of stone formation into regular
patterns at the bottom of a shallow lake.

The subject of thermal convection or generally flow in a two layer system has been studied in much detail with a review of the early work and 
applications to various areas in industry or geophysics given in chapter 6 of \citet{Straughan:2008}. The intense interest in this class of problem
has been driven by the many applications to diverse areas such as heat pipe technology, renewable energy, see e.g. \citet{Straughan:2008},
contaminant dispersal in water ways, \citet{HibiTomigashi:2015}, \citet{Hibi:2020}, or even blood flow in arteries and veins in the human body, 
see e.g. \citet{SharmaYadav:2017}, \citet{Tiwari:2020}, \citet{PonalagusamyManchi:2021}, \citet{WajihahSankar:2021}. Indeed, the last mentioned
article involves 5 layer flow comprising Darcy media, Brinkman media, plasma, core flow, and plug flow.

There are many recent stability analyses of fluid flow in the two layer fluid - porous configuration, see e.g. 
\citet{CarrStraughan:2003}, \citet{Chang:2006},
\citet{HillStraughan:2008,HillStraughan:2009awr}, \citet{Chang:2017}, \citet{Samanta:2020}, \citet{Yin:2020}, \citet{Tsiberkin:2020}. 
Particular analyses
involving the nonlinear theory and bifurcations are given by \citet{Han:2020}, \citet{McCurdy:2019}, \citet{LyuWang:2021},
and \citet{HillStraughan:2009}. In addition, mathematical analysis of the structural stability of the two layer system has been thoroughly
investigated, see e.g. \citet{Li:2018}, \citet{Li:2021}, \citet{Li:2021amo}, \citet{PayneStraughan:1998}.

In a separate development, there has been immense interest in thermal convection in a single layer of saturated porous material but when the porous
skeleton is of double porosity type. By double porosity we mean that the solid skeleton contains pores of a visible size known as macro pores, but
the skeleton itself contains cracks or fissures which give rise to much smaller micropores. Flow in such materials is additionally called
bidisperse or bidispersive. 
The thermal convection problem in a bidisperse porous material was first developed by \citet{NieldKuznetsov:2006} who allowed for different velocities,
pressures and temperature fields in the macro and micro phases and 
a critical review of the topic is given by \citet{GentileStraughan:2020}. There are many recent contributions
driven by the need to understand bidispersive convection in real life applications, cf. \citet{GentileStraughan:2020}, \citet{Straughan:2017},
chapter 13. Analyses of bidispersive thermal convection in isotropic, anisotropic, vertical layer, inclined layer, and with rotation effects are given
by \citet{BaddayHarfash:2021}, \citet{Capone:2020prsa}, \citet{Capone:2020mrc}, \citet{CaponeDeLuca:2020}, \citet{CaponeMassa:2021},
\citet{Capone:2021am}, \citet{Chaloob:2021}, \citet{Falsaperla:2016}, \citet{GentileStraughan:2017,GentileStraughan:2017prsa},
\citet{SaravananVigneshwaran:2020}, \citet{Straughan:2018prsa,Straughan:2019prsa,Straughan:2019abc}, and the structural stability aspect of the system
of equations is considered by \citet{Franchi:2017}.

The object of the current work is to present a model for thermal convection in an incompressible viscous fluid when that fluid overlies a
bidispersive porous medium saturated with the same fluid. We analyse the instability of thermal convection in this system and show there are definite
relations between the onset of convective motion and the respective depths of the fluid and porous layers, and of the properties of the macro and
micro pores. This is the first analysis we have seen of this problem and we believe it has much future application to diverse areas such as
blood flow, heat transfer, and renewable energy.    

\section{Basic equations} 
\label{equations}

We suppose a linearly viscous incompressible fluid is contained in the infinite layer $\mathbb{R}^2\times\{0<z<d\}$ and below this is a bidisperse
porous medium saturated with the same fluid and this occupies the infinite layer $\mathbb{R}^2\times\{-d_m<z<0\}$, with gravity acting in the 
negative $z-$direction.

The equations for the fluid in the layer $\mathbb{R}^2\times\{0<z<d\}$ are then, cf. \citet{Chandrasekhar:1981},
\begin{equation}\label{E:FluidEqs}
\begin{aligned} 
&V_{i,t}+V_jV_{i,j}=-\frac{1}{\rho_0}\,p_{,i}+\nu\Delta V_i+\gamma g k_i T,\\
&V_{i,i}=0,\\
&T_{,t}+V_iT_{,i}=\frac{k_f}{(\rho_0c_p)_f}\,\Delta T,
\end{aligned} 
\end{equation} 
where $V_i({\bf x},t)$ is the velocity field, $T({\bf x},t)$ is the temperature field, $p({\bf x},t)$ is the pressure field, ${\bf x}$ is the spatial
point in the layer, and $t$ is time.
We use indicial notation throughout in conjunction with the Einstein summation convention, so that, for example,
\begin{align*}
V_iT_{,i}\equiv \sum_{i=1}^3V_iT_{,i}&\equiv 
V_1\frac{\partial T}{\partial x_1}
+V_2\frac{\partial T}{\partial x_2}
+V_3\frac{\partial T}{\partial x_3}\\
&\equiv 
U\frac{\partial T}{\partial x}
+V\frac{\partial T}{\partial y}
+W\frac{\partial T}{\partial z}\,,
\end{align*}
where ${\bf V}=(V_1,V_2,V_3)\equiv (U,V,W)$. In equations \eqref{E:FluidEqs} $\gamma,g,\rho_0,k_f,\nu$ and $c_p$ are the thermal expansion 
coefficient of the fluid, gravity, reference density, thermal conductivity of the fluid, kinematic viscosity of the fluid, and specific heat at
constant pressure of the fluid. The vector ${\bf k}=(0,0,1)$ and $\Delta$ is the Laplacian
\begin{equation*}
\Delta=\frac{\partial^2}{\partial x^2}
+\frac{\partial^2}{\partial y^2}
+\frac{\partial^2}{\partial z^2}\,.
\end{equation*}

For an isotropic bidisperse porous material we suppose the macro porosity is $\phi$, the micro porosity is $\epsilon$. If we denote
$(U^f_i,p^f)$ to be the pore averaged velocity and pressure in the macropores and 
$(U^p_i,p^p)$ to be the pore averaged velocity and pressure in the micropores, then the governing equations of flow in the bidisperse porous 
medium may be written, cf. \citet{GentileStraughan:2017}, \citet{Straughan:2018prsa},
\begin{equation}\label{E:PorEqs}
\begin{aligned} 
&-\frac{\mu}{K_f}\,U^f_i-\zeta(U^f_i-U^p_i)-p^f_{,i}+\rho_0\gamma k_i g T^m=0,\\
&U^f_{i,i}=0,\\
&-\frac{\mu}{K_p}\,U^p_i-\zeta(U^p_i-U^f_i)-p^p_{,i}+\rho_0\gamma k_i g T^m=0,\\
&U^p_{i,i}=0,\\
&(\rho_0c)_mT^m_{,t}+(\rho_0c)_f(U^f_i+U^p_i)T^m_{,i}=k_m\Delta T^m,
\end{aligned} 
\end{equation} 
where $T^m({\bf x},t)$ is the temperature field of the fluid in the bidisperse porous medium.

Equations \eqref{E:FluidEqs} hold on the domain $\{(x,y)\in\mathbb{R}^2\}\times\{z\in(0,d)\}\times\{t>0\}$ while
\eqref{E:PorEqs} hold on the domain 
$\{(x,y)\in\mathbb{R}^2\}\times\{z\in(-d_m,0)\}\times\{t>0\}$.
Equations \eqref{E:PorEqs} assume Darcy's law holds and a Boussinesq approximation is employed. The variable $\mu$ is the dynamic viscosity of the
fluid, $K_f$ and $K_p$ are the macro and micro permeabilities, $\zeta$ is an interaction coefficient which represents the momentum transfer 
between the macro and micro phases, and $(\rho_0c)_m$, $k_m$ are given by
\begin{equation*}
(\rho_0c)_m=(1-\phi)(1-\epsilon)(\rho_0c)_s+\phi(\rho_0c)_f+\epsilon(1-\phi)(\rho_0c)_p\,,
\end{equation*}
and 
\begin{equation*}
k_m=(1-\phi)(1-\epsilon)k_s+\phi k_f+\epsilon(1-\phi) k_p\,,
\end{equation*}
where $s,f$ and $p$ denote values in the solid skeleton, the fluid in the macropores, and the fluid in the micropores.

The boundary conditions at the top and bottom of the layer, $z=d$ and $z=-d_m$ are specified as follows
\begin{equation}\label{E:BCs1}
\begin{aligned}
& V_i=0,\quad T=T_U,\qquad {\rm on}\,\,z=d,\\
& U^f_3=0,\quad U^p_3=0,\quad T^m=T_L,\qquad {\rm on}\,\,z=-d_m\,,
\end{aligned}
\end{equation}
where $T_U,T_L$ are constants with $T_L>T_U>0$. Under these conditions \eqref{E:FluidEqs} and \eqref{E:PorEqs} admit a steady conduction 
solution of form
\begin{equation}\label{E:SSoln}
\begin{aligned}
&{\bar U}^f_i=0,\qquad{\bar U}^p_i=0,\qquad {\bar V}_i=0,\\
&{\bar T}=T_0-(T_0-T_U)\,\frac{z}{d}\,,\qquad 0\le z\le d,\\
&{\bar T}^m=T_0-(T_L-T_0)\,\frac{z}{d_m}\,,\qquad -d_m\le z\le 0,
\end{aligned}
\end{equation}
where we have employed the fact that the temperature is continuous across the interface $z=0$. To determine the constant $T_0$ we require the 
heat flux to be continuous across the interface $z=0$ and then
\begin{equation*}
k_m\,\frac{d{\bar T}^m}{dz}=k_f\frac{d{\bar T}}{dz}\,,\qquad {\rm at}\,\,z=0.
\end{equation*}
This yields $T_0$ as 
\begin{equation}\label{E:T0val}
T_0=\frac{k_fT_Ud_m+k_mT_Ld}{dk_m+d_mk_f}\,.
\end{equation}

\section{Perturbation equations} 
\label{pert}

As our goal is to study instability of the base solution \eqref{E:SSoln} we now introduce perturbations to the variables
$V_i,T,p^f,U^f_i,U^p_i,T^m,p^p$
as
$u_i,\theta,\pi^f,u_i^f,u_i^p,\theta^m$
and
$\pi^p$.
We derive the perturbation equations for these variables from equations \eqref{E:FluidEqs} and \eqref{E:PorEqs}. However, it is convenient to present these
in non-dimensional form with length scales
$d,d_m$
in the fluid and bidisperse porous layers, and with corresponding velocity scales
$U$ and $U^m$,
although we here select
$U=U^m$.
The time scales are 
$\mathcal{T}$ and $\mathcal{T}^m$
where
$\mathcal{T}=d^2/\nu$,
with
$\nu=\mu/\rho_0$
and we pick
$\mathcal{T}=\mathcal{T}^m$.
The pressure scale is
$P=\mu U/d$.
The temperature scales are
\begin{equation*}
T^{\sharp}=\frac{U(T_0-T_U)d(\rho_0c_p)_f}{k_f}
\end{equation*}
and
\begin{equation*}
T^{\sharp}_m=U^m(T_L-T_0)\frac{d_m(\rho_0c_p)_f}{k_m}
\end{equation*}
and thus one finds
\begin{equation*}
\frac{T^{\sharp}}{T^{\sharp}_m}=\Bigl(\frac{k_m}{k_f}\Bigr)^2{\hat d}^2\,,\qquad
\frac{T_0-T_U}{T_L-T_0}=\frac{k_m}{k_f}\,{\hat d}
\end{equation*}
where
${\hat d}=d/d_m$.
It is convenient to introduce the notation
\begin{equation*}
{\hat k}=\frac{k_f}{k_m}\,,\qquad{\hat\kappa}=\frac{(\rho_0c)_m}{(\rho_0c)_f}\,{\hat k}\,,
\end{equation*}
and to define the Prandtl number, $Pr$, and the porous Prandtl number, $Pr_m$, by
\begin{equation*}
Pr=\frac{\nu}{\kappa}=\frac{\mu}{\rho_0}\,\frac{(\rho_0c_p)_f}{k_f}\,,\qquad
Pr_m=\frac{\mu}{\rho_0}\,\frac{(\rho_0c)_m}{k_m}
\end{equation*}
from which one may show
$Pr_m={\hat\kappa}Pr$.

The fluid and porous Rayleigh numbers $Ra$ and $Ra^m$ are defined by
\begin{equation}\label{E:RaDef}
Ra=\frac{\gamma gd^4}{\kappa_f\nu}\,\frac{(T_0-T_U)}{d}\,,
\end{equation}
where
$\kappa_f=k_f/(\rho_0c_p)_f$, 
and
\begin{equation}\label{E:RamDef}
Ra_m=\gamma gK^f\frac{(T_L-T_0)}{d_m}\,\frac{d_m^2}{[k_m/(\rho_0c)_f]\nu}\,,
\end{equation}
from which one shows
\begin{equation*}
Ra=\Bigl(\frac{{\hat d}}{{\hat k}}\Bigr)^2\,\frac{Ra_m}{Da}
\end{equation*}
where $Da$ is the Darcy number defined here as
\begin{equation*}
Da=\frac{K_f}{d^2}\,.
\end{equation*}
The relative permeability $K_r$ is defined by $K_r=K_f/K_p$, and another useful non-dimensional variable is
$\delta=\sqrt{K_p}/d_m$,
from which we may see that
$Da=K_r\delta^2/{\hat d}^2$.

With the above non-dimensionalization one may show that the linearized fluid perturbation equations have form
\begin{equation}\label{E:FluidPert}
\begin{aligned}
&u_{i,t}=-\pi_{,i}+\Delta u_i+Rak_i\theta,\\
&u_{i,i}=0,\\
&Pr\theta_{,t}=w+\Delta\theta
\end{aligned}
\end{equation}
where $w=u_3$ and these equations hold on $\mathbb{R}^2\times\{z\in(0,1)\}\times\{t>0\}$
while the linearized bidispersive porous media equations have form
\begin{equation}\label{E:PorPert}
\begin{aligned}
&-u^f_i-\xi(u_i^f-u_i^p)-\pi^f_{,i}+Ra^mk_i\theta^m=0,\\
&u^f_{i,i}=0,\\
&-K_ru^p_i-\xi(u_i^p-u_i^f)-\pi^p_{,i}+Ra^mk_i\theta^m=0,\\
&u^p_{i,i}=0,\\
&\frac{Pr^m}{{\hat d}^2}\,\theta^m_{,t}=w^f+w^p+\Delta\theta_m
\end{aligned}
\end{equation}
where
$w^f=u_3^f,w^p=u_3^p$ and equations \eqref{E:PorPert} hold on 
$\mathcal{R}^2\times\{z_m\in(-1,0)\}\times\{t>0\}$. The non-dimensional momentum transfer interaction coefficient $\xi$ is defined by
$\xi=\zeta K_f/\mu$.

Equations \eqref{E:FluidPert} and \eqref{E:PorPert} when reduced to the thermal convection instability problem essentially represent a twelth order system
of equations. We thus require twelve boundary conditions. In this work we employ a normal mode instability analysis and it is sufficient to require the
following conditions
\begin{equation*}
w=w^{\prime}=\theta=0,\qquad {\rm on}\,\,z=1,
\end{equation*}
where $w^{\prime}=\partial w/\partial z$. This corresponds to a fixed upper surface. Also,
\begin{equation*}
w^f=w^p=\theta^m=0,\qquad {\rm on}\,\,z=-1.
\end{equation*}
The remaining six boundary conditions come from considerations on the interface $z=0$.
At the micoscopic level the velocity $W$ in the fluid should be continuous across the interface with the actual velocity $W^f$ in a macro pore
and with the actual velocity $W^p$ in a micro pore. In dimensional form this requires at $z=0$,
\begin{equation*}
W=\frac{W^f}{\phi},\qquad{\rm and}\qquad W=\frac{W^p}{\epsilon(1-\phi)}\,,
\end{equation*}
where $w^f$ and $w^p$ are the pore averaged values. Likewise the dimensional temperatures are continuous so
\begin{equation*}
\theta=\theta_m,\qquad{\rm at}\qquad z=0.
\end{equation*}
In addition the normal component of heat flux ${\bf q}\cdot{\bf n}$ is continuous across $z=0$. Two further conditions are needed and these arise by
requiring continuity of normal stress, and by appealing to a combination of appropriate forms of the experimentally verified \citet{BeaversJoseph:1967}
condition. Details of these boundary conditions are amplified below.

\section{Instability analysis} 
\label{instab}

The next step is to remove the pressure terms $\pi,\pi^f$ and $\pi^p$ from \eqref{E:FluidPert} and \eqref{E:PorPert}
and this we do by taking curl curl of equations \eqref{E:FluidPert}$_1$ and \eqref{E:PorPert}$_{1,3}$
and retaining the third component. This results in the 
system of equations
\begin{equation}\label{E:Lin1}
\begin{aligned}
&\sigma\Delta w=\Delta^2w+Ra\Delta^*\theta,\\
&\sigma Pr\theta=w+\Delta\theta,
\end{aligned}
\end{equation}
and
\begin{equation}\label{E:Lin2}
\begin{aligned}
&(1+\xi)\Delta w^f-\xi\Delta w^p-Ra^m\Delta^*\theta^m=0,\\
&(K_r+\xi)\Delta w^p-\xi\Delta w^f-Ra^m\Delta^*\theta^m=0,\\
&\frac{\sigma_m Pr_m}{{\hat d}^2}\,\theta^m=w^f+w^p+\Delta\theta^m,
\end{aligned}
\end{equation}
where we have represented time by $e^{\sigma t}$ in the fluid equations and by $\exp(\sigma_mt)$ in the bidispersive porous equations. The symbol 
$\Delta^*$ is the horizontal Laplacian. Equations \eqref{E:Lin1} hold on 
$\mathbb{R}^2\times(0,1)$ 
while \eqref{E:Lin2} hold on
$\mathbb{R}^2\times(-1,0)$. 

We next solve \eqref{E:Lin2}$_{1,2}$ in terms of $\Delta w^f$ and $\Delta w^p$. We represent $w$ and $\theta$ as
$w=W(z)h(x,y)$,
$\theta=\Theta(z)h(x,y)$,
and
$w^f=W^f(z)h^m(x,y)$,
$w^p=W^p(z)h^m(x,y)$,
$\theta_m=\Theta_m(z)h^m(x,y)$,
where $h$ and $h_m$ are plan forms which tile the plane, cf. \citet{Chandrasekhar:1981}, pp. 43-52,
and are typical of the hexagonal convection cell forms found in  real life. The functions $h$ and $h_m$ satisfy the relations $\Delta^*h=-a^2h$
and
$\Delta^*h_m=-a_m^2h_m$,
for wavenumbers $a$ and $a_m$.
We reduce \eqref{E:Lin1}$_1$ to two second order equations by setting $\Delta w=\chi$, and then we arrive at the following coupled system of equations
to solve for the growth rate (eigenvalues)
$\sigma,\sigma_m$, 
\begin{equation}\label{E:Lin3}
\begin{aligned}
&(D^2-a^2)W-\chi=0,\\
&(D^2-a^2)\chi-Ra\,a^2\Theta=\sigma\chi,\\
&(D^2-a^2)\Theta+W=Pr\,\sigma\,\Theta,
\end{aligned}
\end{equation}
on $z\in(0,1)$, where $D=d/dz$, and
\begin{equation}\label{E:Lin4}
\begin{aligned}
&(D^2-a^2_m)W^f+Ra_m\,a_m^2\,\frac{(K_r+2\xi)}{(K_r+\xi+\xi K_r)}\,\Theta_m=0,\\
&(D^2-a^2_m)W^p+Ra_m\,a_m^2\,\frac{(1+2\xi)}{(K_r+\xi+\xi K_r)}\,\Theta_m=0,\\
&(D^2-a^2_m)\Theta_m+W^f+W^p=\frac{Pr_m}{{\hat d}^2}\,\sigma_m\,\Theta_m,
\end{aligned}
\end{equation}
on $z_m\in(-1,0)$.

The non-dimensional boundary conditions are 
\begin{equation}\label{E:BCs1A}
\begin{aligned}
&W=W^{\prime}=\Theta=0,\qquad z=1,\\
&W^f=W^p=\Theta^m=0,\qquad z=-1,
\end{aligned}
\end{equation}
together with the non-dimensional interface conditions
\begin{equation}\label{E:BCs2}
\begin{aligned}
&W=\frac{W^p}{\epsilon(1-\phi)}\,,\qquad\frac{W=W^f}{\phi}\,,\qquad z=0,\\
&\theta_m=\theta\Bigl(\frac{k_m}{k_f}\Bigr)^2\,{\hat d}^2,\qquad
\frac{d\theta_m}{dz_m}=\frac{d\theta}{dz}\,{\hat d}\,\frac{k_m}{k_f}\,,\qquad z=0,
\end{aligned}
\end{equation}
where the latter two arise due to continuity of temperature, and continuity of heat flux.

For the remaining interface conditions we argue as follows.

In terms of the fluid and pore averaged velocities at the microscopic level, the \citet{BeaversJoseph:1967} condition may be applied separately to the macro
and micro components to yield in dimensionless form
\begin{equation}\label{E:BJf}
\frac{1}{d}\,\frac{\partial u_{\beta}}{\partial z}=\frac{\alpha_1}{\sqrt{K_f}}\,(u_{\beta}-u^f_{\beta})\,,\qquad\beta=1,2,
\end{equation}
and
\begin{equation}\label{E:BJp}
\frac{1}{d}\,\frac{\partial u_{\beta}}{\partial z}=\frac{\alpha_2}{\sqrt{K_p}}\,(u_{\beta}-u^p_{\beta})\,,\qquad\beta=1,2,
\end{equation}
where $\alpha_1$ and $\alpha_2$
are experimentally determined constants. For a single porosity material \citet{BeaversJoseph:1967} write that ``{\sl $\alpha$ is a dimensionless
quantity depending on the material parameters which characterize the structure of permeable material within the boundary region}".
Our first approach is to take an averaged value of each of the conditions \eqref{E:BJf} and \eqref{E:BJp} and propose at the interface $z=0$,
\begin{equation}\label{E:BJ1}
\frac{\partial u_{\beta}}{\partial z}=(A_1+A_2)u_{\beta}-A_1u^f_{\beta}-A_2u^p_{\beta}\,,\qquad\beta=1,2,
\end{equation}
where
\begin{equation}\label{E:BJA1A2}
A_1=\frac{\alpha_1d}{2\sqrt{K_f}}\,\qquad A_2=\frac{\alpha_2d}{2\sqrt{K_p}}\,.
\end{equation}
We next differentiate \eqref{E:BJ1} and employ the incompressibility conditions to derive the interface condition
\begin{equation}\label{E:BJ11}
w_{zz}=(A_1+A_2)w_z-A_1{\hat d}w^f_{z_m}-A_2{\hat d}w^p_{z_m}
\end{equation}
where the derivatives are with respect to $z,z_m$ non-dimensional, $0\le z\le 1$, $-1\le z_m\le 0$.

We also consider an alternative procedure whereby we combine the \citet{BeaversJoseph:1967} conditions \eqref{E:BJf} and \eqref{E:BJp} in a manner which
reflects the macro and micro contributions.
Thus, define the constants
$c_1$ and $c_2$ by
\begin{equation*}
c_1=\frac{\phi}{\phi+\epsilon(1-\phi)}\,,\qquad c_2=\frac{\epsilon(1-\phi)}{\phi+\epsilon(1-\phi)}\,.
\end{equation*}
We now produce an equation of form \eqref{E:BJ1} but where now 
\begin{equation}\label{E:A1A2def}
A_1=c_1\alpha_1d/\sqrt{K_f}\qquad
A_2=c_2\alpha_2d/\sqrt{K_p}\,.
\end{equation}
We thus derive an equation of form \eqref{E:BJ11} but with $A_1$ and $A_2$
as given here. The differences in employing \eqref{E:BJ11} and the pore weighted version just discussed are considered in the numerical results section.
For ease in understanding the numerical results where the effect of parameter variation upon the solution is considered we specifically write the two
versions of \eqref{E:BJ11} here as, equal splitting,
\begin{equation}\label{E:BJes}
w_{zz}=\Bigl(\frac{\alpha_1d}{2\sqrt{K_f}}+\frac{\alpha_2d}{2\sqrt{K_p}}\Bigr)w_z-\frac{\alpha_1d}{2\sqrt{K_f}}\,{\hat d}\,w^f_{z_m}
-\frac{\alpha_2d}{2\sqrt{K_p}}\,{\hat d}\,w^p_{z_m}
\end{equation}
and pore weighted
\begin{equation}\label{E:BJpw}
w_{zz}=\Bigl(\frac{c_1\alpha_1d}{\sqrt{K_f}}+\frac{c_2\alpha_2d}{\sqrt{K_p}}\Bigr)w_z-\frac{c_1\alpha_1d}{\sqrt{K_f}}\,{\hat d}\,w^f_{z_m}
-\frac{c_2\alpha_2d}{\sqrt{K_p}}\,{\hat d}\,w^p_{z_m}\,.
\end{equation}

At the microscopic level, continuity of normal stress at the interface $z=0$ requires
\begin{equation*}
\pi^f=\pi-2\mu w_z,\qquad {\rm for}\,\,{\rm a}\,\,{\rm macropore},
\end{equation*}
and
\begin{equation*}
\pi^p=\pi-2\mu w_z,\qquad {\rm for}\,\,{\rm a}\,\,{\rm micropore}.
\end{equation*}
Our first continuity of normal stress interface condition is
\begin{equation}\label{E:CtyNS}
\frac{\pi^f+\pi^p}{2}=\pi-2\mu w_z,\qquad {\rm on}\,\,z=0.
\end{equation}
Equation \eqref{E:CtyNS} is differentiated with respect to $x_{\alpha}$, $\alpha=1,2$,
and one then employs the differential equations \eqref{E:FluidPert}$_{1,2}$ and \eqref{E:PorPert}$_{1-4}$ in the forms
\begin{equation}\label{E:Int}
\begin{aligned}
&\pi_{,\alpha}=\Delta u_{\alpha}-\sigma u_{\alpha}\,,\qquad u_{\alpha,\alpha}+w_{,z}=0,\\
&\pi^f_{,\alpha}=-u^f_{\alpha}-\xi(u^f_{\alpha}-u^p_{\alpha})\,,\qquad u^f_{\alpha,\alpha}+w^f_{,z}=0,\\
&\pi^p_{,\alpha}=-K_r u^p_{\alpha}-\xi(u^p_{\alpha}-u^f_{\alpha})\,,\qquad u^p_{\alpha,\alpha}+w^p_{,z}=0,
\end{aligned}
\end{equation}
to eliminate the pressure terms. The differentiated form of \eqref{E:CtyNS} is rewritten using \eqref{E:Int} as
\begin{equation}\label{E:Int2}
\frac{1}{2}\bigl(-u^f_{\alpha}-\xi[u^f_{\alpha}-u^p_{\alpha}]-K_ru^p_{\alpha}-\xi[u^p_{\alpha}-u^f_{\alpha}]\bigr)
=\Delta u_{\alpha}-\sigma u_{\alpha}-2\mu w_{,z\alpha}\,,
\end{equation}
where $\alpha=1,2$.
Note that in this case the $\xi$
terms disappear.
Now differentiate \eqref{E:Int2} for $\alpha=1$ with respect to $x$ and for $\alpha=2$ with respect to $y$. This yields after summation to
\begin{equation*}
\frac{1}{2}(-u^f_{\alpha,\alpha}-K_r u^p_{\alpha,\alpha})=\Delta u_{\alpha,\alpha}-\sigma u_{\alpha,\alpha}-2\mu w_{,z\alpha\alpha}\,.
\end{equation*}
Then use the incompressibility conditions to find
\begin{equation}\label{E:Int3}
\frac{1}{2}(w^f_{,z}+K_rw^p_{,z})=-\Delta w_{,z}+\sigma w_{,z}-2\mu\Delta^*w_{,z}
\end{equation}
where $\Delta^*=\partial^2/\partial x^2+\partial^2/\partial y^2$.
Equation \eqref{E:Int}$_3$ allows one to determine the following non-dimensional interface condition
\begin{equation}\label{E:CtyNS2}
\frac{1}{2}(D_mw^f+KrD_mw^p)=\frac{Da}{{\hat d}}\,(\sigma Dw-D^3w-3\Delta^*Dw),
\end{equation}
where $D=d/dz,z\in(0,1)$,
$D_m=d/dz_m,z_m\in(-1,0)$.

Alternatively, one may employ a weighted form of \eqref{E:CtyNS} where we write
\begin{equation*}
c_1\pi^f+c_2\pi^p=\pi-2\mu w_z\,.
\end{equation*}
This then leads to a weighted form of \eqref{E:CtyNS2}. In deriving the weighted form the interaction terms involving $\xi$ do not vanish. The precise forms
are given in the next section.

Thus, the complete set of boundary conditions are \eqref{E:BCs1A}, \eqref{E:BCs2}, together with \eqref{E:BJ11} and \eqref{E:CtyNS2}, or the 
weighted equivalents of these latter two equations.

\section{Numerical method} 
\label{nummeth}

We solve equations \eqref{E:Lin3} and \eqref{E:Lin4} by a Chebyshev tau method, cf. \citet{Dongarra:1996}, coupled with the $QZ$ algorithm for a generalized 
matrix eigenvalue problem, cf. \citet{MolerStewart:1971}.
Equations \eqref{E:Lin3} are transformed into the Chebyshev domain $(-1,1)$ and equations \eqref{E:Lin4} are likewise transformed into the same domain with
the interface now being $z=-1$. The variables 
$W,\chi,\Theta,W^f,W^p$ and $\Theta^m$ are written as finite series of Chebyshev polynomials, e.g.
\begin{equation*}  
W=\sum_{i=0}^N W_iT_i(z)\,,
\end{equation*}  
for Fourier coefficients $W_i$. This yields a block matrix generalized eigenvalue problem of form
$A{\bf x}=\sigma B{\bf x}$
for $6N\times 6N$
matrices $A,B$ with $B$ singular, where
\begin{equation*}  
{\bf x}=({\tilde W},{\tilde\chi},{\tilde\Theta},{\tilde W^f},{\tilde W^p},{\tilde\Theta^m})
\end{equation*}  
${\tilde W}$ etc., being the truncated versions of $W$, etc., i.e.
\begin{equation*}  
{\tilde W}=(W_0,\cdots,W_N),\cdots,{\tilde\Theta^m}=(\Theta^m_0,\cdots,\Theta^m_N).
\end{equation*}  
The boundary conditions are likewise expanded in Chebyshev polynomials and added as rows of the 
matrices $A$ and $B$ via a similar procedure to that explained in \citet{Dongarra:1996}.

The complete set of boundary conditions on $z=1$ or $z=-1$ are now
\begin{equation}\label{E:BCsCD}
\begin{aligned}
&W=0,\,\,DW=0,\,\,\Theta=0,\,\,W^f=0,\,\,W^p=0,\,\,\Theta_m=0,\quad z=1,\\
&W=\frac{W^f}{\phi}\,,\quad W=\frac{W^p}{\epsilon(1-\phi)}\,,\quad z=-1,\\
&\Theta_m=\Theta\,\frac{{\hat d}^2}{{\hat k}^2}\,,\quad D_m\Theta_m+\frac{{\hat d}}{{\hat k}}\,D\Theta=0,\quad z=-1,\\
&A+a^2W-2(A_1+A_2)DW\\
&\qquad -2A_2{\hat d}D_mW^p-2A_1{\hat d}D_mW^f=0,\quad z=-1,
\end{aligned}
\end{equation}
and
\begin{equation}\label{E:BCcns}
2DA-4a^2DW-\frac{{\hat d}}{Da}\,D_mW^f-\frac{K_r{\hat d}}{Da}\,D_mW^p=2\sigma DW,\quad z=-1.
\end{equation}
This is for the case \eqref{E:CtyNS} and \eqref{E:BJ11} with $A_1$ and $A_2$ given by \eqref{E:BJA1A2}. When the weighted version of the Beavers - Joseph
and continuity of normal stress conditions are employed then \eqref{E:BCsCD}$_3$ holds but with $A_1$ and $A_2$ given by \eqref{E:A1A2def}.
However, in the weighted case \eqref{E:BCcns} should be replaced by
\begin{equation}\label{E:BCWtd}
\begin{aligned}
2DA&-4a^2DW-\frac{2{\hat d}}{Da}\,[c_1+\xi(c_1-c_2)]D_mW^f\\
&-\frac{2{\hat d}}{Da}[K_rc_2+\xi(c_2-c_1)]D_mW^p=2\sigma DW,
\end{aligned}
\end{equation}
where it is to be observed that the interaction terms involving $\xi$ are present.

\section{Numerical results} 
\label{num}

This section reports on numercial results for the critical Rayleigh number and critical wavenumber for the equations for thermal convection in a 
linearly viscous fluid overlying a bidisperse porous material. We choose parameter values appropriate to water being the working fluid and a bidisperse
porous material being based upon a glass bead skeleton.

As this is the first calculation for this problem we restrict attention to $\alpha_1=\alpha_2=\alpha$ for the Beavers - Joseph constant. Since there are many
parameters we believe this is justified. For numerical values of the many parameters we refer to \citet{GentileStraughan:2020} who employed tabulated
experimental values of \citet{Hooman:2015}, \citet{ImaniHooman:2017}, \citet{Chen:1991}, \citet{Rees:2009,Rees:2010} and \citet{Nield:2000}.

\citet{BeaversJoseph:1967} reported values of $\alpha$ for a single porosity material in the range 0.1 to 4. These values are for a granular aloxite
material and for man made porous foams. In this work we concentrate on more granular materials and choose $\alpha$ values close to 0.1. The values reported
in \citet{GentileStraughan:2020} suggest we take relative permeabilities of
\begin{equation*}
K_r=25,151.7,263.16
\end{equation*}
and values for the non-dimensional momentum transfer coefficient in the group
\begin{equation*}
\xi=1.515\times 10^{-2},2.347\times 10^{-2},2.987\times 10^{-2},0.1316.
\end{equation*}

We here choose to investigate the behaviour of the critical $Ra_m$ and $a_m$ values upon 
$K_r,\alpha,\xi,\phi,\epsilon$, and upon the equal splitting interface conditions of Beavers and Joseph
and continuity of normal stress together with the analogous pore weighted interface conditions.

For core values using glass beads and water the thermal conductivities, densities and specific heats are taken from the internet version of
Engineering Toolbox to yield $k_m,k_f,(\rho c)_m$ and $(\rho c)_f$. We also employ the parameter ranges in \citet{GentileStraughan:2020} to find
values for $\xi,K_r$ and $\delta$.
In this way we obtain
$Pr=6,Pr_m=0.75828,$ $Da=0.161278\times 10^{-2}$, $\delta=0.3279\times 10^{-2},$ ${\hat k}=0.16736$, $\phi=0.3,\epsilon=0.3$, $\alpha=0.1$,
$\xi=0.02987$, ${\hat\kappa}=0.12638$, $K_r=25$, although specific values will be varied at appropriate points in our discussion. 
The nature of the onset of convective motion in all cases depends on ${\hat d}$, the depth of fluid layer to depth of porous layer.
There is a critical value of ${\hat d}$ such that when ${\hat d}$ is below this value then convective motion commences in the porous layer
whereas such motion is initiated in the fluid layer when ${\hat d}$ is above the critical value. The critical ${\hat d}$ value also depends
strongly on the other parameters in the problem and this variation is examined in detail here.

\begin{figure}
\begin{center}
\includegraphics[width=\linewidth]{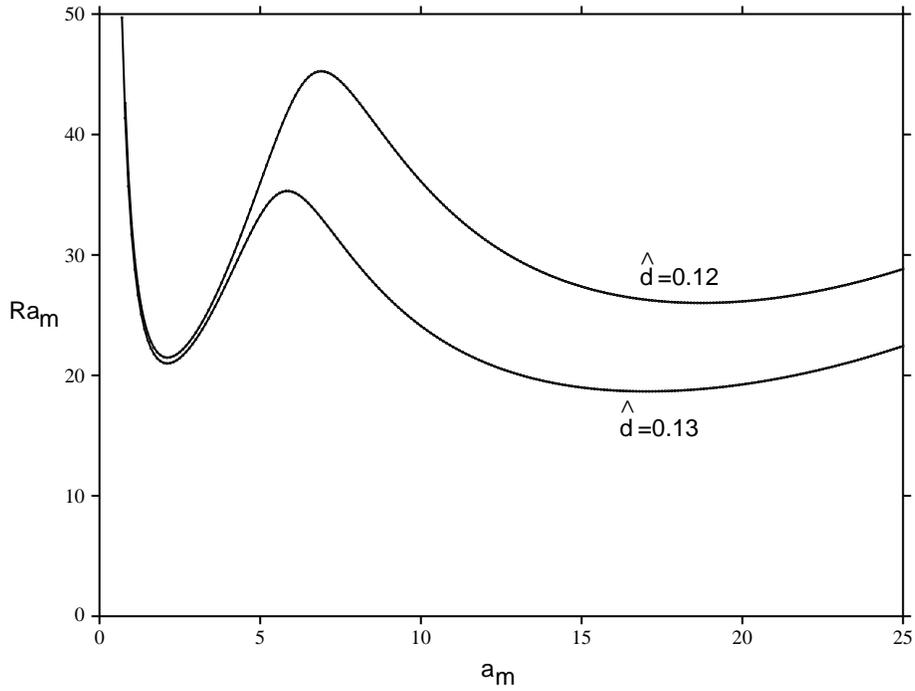}
\caption{
Graph of
$Ra_m$ vs. $a_m$.
Here,
$\delta=0.003279$,
${\hat k}=0.16736$,
$\phi=0.3$,
$\epsilon=0.3$,
$\alpha=0.1$,
$\xi=0.02987$,
${\hat\kappa}=0.12638$,
$Pr=6$,
$K_r=25$.
The minimum values on the 
${\hat d}=0.12$
curve are at 
$a_m=2.1,\,Ra_m=21.474$
and 
$a_m=18.7,\,Ra_m=26.015,$
on the
${\hat d}=0.13$
curve they are
$a_m=2.1,\,Ra_m=20.991$
and
$a_m=17.0,\,Ra_m=18.671.$
The instability when  
${\hat d}=0.12$
initiates in the porous medium, whereas when
${\hat d}=0.13$
it initiates in the fluid.
}
\label{fig:fig1}
\end{center}
\end{figure}

\begin{figure}
\begin{center}
\includegraphics[width=\linewidth]{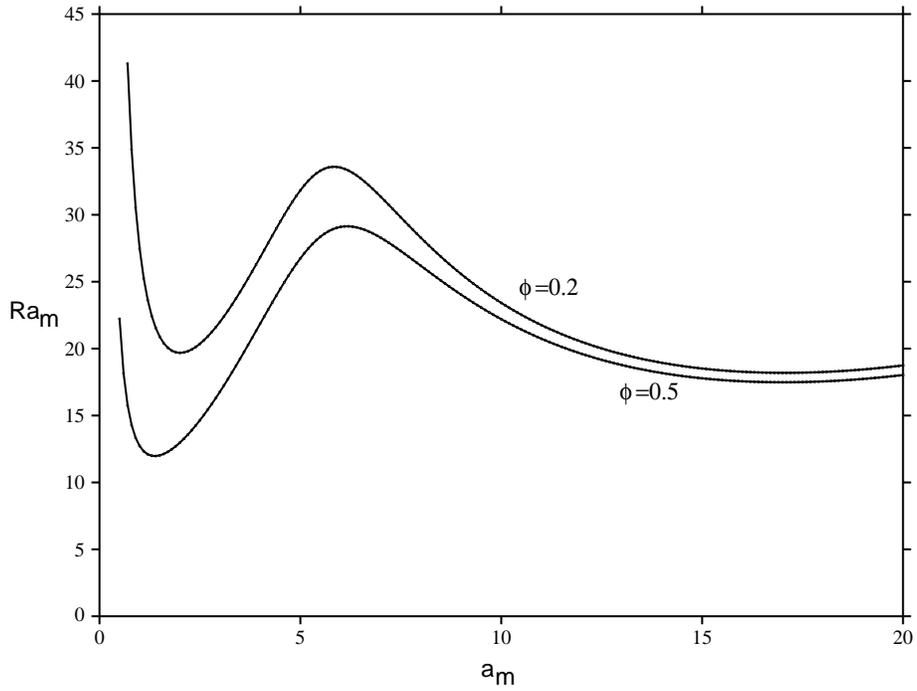}
\caption{
Graph of
$Ra_m$ vs. $a_m$.
Here,
$\delta=0.003279$,
${\hat k}=0.16736$,
${\hat d}=0.13$,
$\epsilon=0.2$,
$\alpha=0.1$,
$\xi=0.02987$,
${\hat\kappa}=0.12638$,
$Pr=6$,
$K_r=25$.
The minimum values on the 
$\phi=0.2$
curve are at 
$a_m=2.0,\,Ra_m=19.683$
and 
$a_m=17.1,\,Ra_m=18.191,$
on the
$\phi=0.5$
curve they are
$a_m=1.4,\,Ra_m=11.973$
and
$a_m=17.0,\,Ra_m=17.486.$
The instability when  
$\phi=0.5$
initiates in the porous medium, whereas when
$\phi=0.2$
it initiates in the fluid.
}
\label{fig:fig2}
\end{center}
\end{figure}

The bimodal behaviour of the neutral curves is displayed in figures \ref{fig:fig1} and \ref{fig:fig2}. Figure \ref{fig:fig1} shows that
when ${\hat d}=0.12$ the lowest minimum for $Ra_m$ is at $a_m=2.1$ and this corresponds to convection initiating in the porous layer, whereas
when ${\hat d}=0.13$ the lowest minimum for $Ra_m$ is at $a_m=17.0$ and this corresponds to convective motion initiating in the fluid layer. From table 
\ref{Ta:Ta1} we see that the critical value for ${\hat d}$ corresponding to figure \ref{fig:fig1} is when ${\hat d}_{crit}\in(0.1260,0.1261)$.
Observe that from table \ref{Ta:Ta1} the wavenumber jumps from $a_m=2.1$ to $a_m=17.7$ as the initiation of convection switches from the porous layer 
to the fluid layer. Since $a_m$ is inversely proportional to the aspect ratio of the convection cell this means that for a non-dimensional depth of 
porous layer of 1 the width of the cell changes from $2\pi/2.1$ to $2\pi/17.7$, i.e. from 2.99 to 0.355, or from wide cells in the porous layer
to cells 8.42 times smaller in the fluid layer. Of course, the aspect ratio also depends on other parameters in the bidispersive porous medium.
Figure \ref{fig:fig2} also displays a switch from convection in the porous layer to convection in the fluid layer, but now when ${\hat d}$
is fixed. In this case the switch of convection is due to the macro porosity changing. 

It is noticeable that the minima associated to the fluid are widely separated in figure \ref{fig:fig1} whereas in \ref{fig:fig2} it is the minima 
associated to the porous medium which display the greater variation. Since figure \ref{fig:fig2}
is displaying changes in the porosity the larger variation in the ``porous minimum" is to be expected.

Table \ref{Ta:Ta1} displays how the critical values of the porous Rayleigh number $Ra_m$ and the critical values of the porous wavenumber $a_m$ change as
$K_r$ is varied. Since we are effectively fixing $K^f$ in our computations changing $K_r$ corresponds to changing $1/K^p$. As $K_r$ increases in
table \ref{Ta:Ta1} from values of 1.5 through to 263.16 we see that $Ra_m$ increases, both for the porous and fluid minima. This corresponds to the 
layered system becoming less easy to convect as $\Delta T=T_L-T_U$ increases. Since $K_r$ increasing corresponds to $K^p$ decreasing this means it is more 
difficult for the fluid to move in the micro pores and so we expect the system to be more stable. It is worth observing that in all our computations we have
found that at criticality the growth rate $\sigma$ is real.

From table \ref{Ta:Ta1} the porous wavenumber $a_m$ increases by a factor of approximately 2.5 over the range of variation of $K_r$ but the wavenumber
corresponding to the fluid decreases from 24.4 or 26.1 to 9.3 or 8.3, respectively, depending on the value of ${\hat d}$. Thus, at the onset of convection when the 
motion is initiated in the porous layer increasing $K_r$ decreases the cell aspect ratio whereas the opposite is true when the convective motion is initiated
in the fluid layer. The quantitative value of this effect does demonstrate that the presence of the bidisperse layer is significant in both cell aspect 
ratio and whether convection will occur.

Table \ref{Ta:Ta2} displays how $Ra_m$ and $a_m$ change at criticality with variation in the Beavers - Joseph interface parameter. We find that increasing
$\alpha$ decreases the value of $Ra_m$ and also that of $a_m$. Since the coefficient $\alpha$ multiplies $u_{\alpha}-u^{f,p}_{\alpha}$ increasing
its value has the effect of increasing the shear flow term $\partial u_{\alpha}/\partial z$ which is making the system more stable and increasing the
convection cell aspect ratio.

Table \ref{Ta:Ta3} displays the effect of changing the momentum transfer coefficient $\xi$ upon $Ra_m$ and $a_m$. It is seen that a relative change of
over 4 in $\xi$ reduces both $Ra_m$ and $a_m$, although the change is relatively small.

Table \ref{Ta:Ta4} demonstrates how the critical values of $Ra_m$ and $a_m$ are affected by changing the macro porosity $\phi$ and the micro porosity 
$\epsilon$. For a fixed value of $\phi$ increasing $\epsilon$ leads to an increase in $Ra_m$ and $a_m$ for the convection initiation in the porous 
medium, while the values associated to initiation in the fluid part of the layer show little variation. This indicates that increasing $\epsilon$ for fixed $\phi$
helps to stabilize the fluid layer. This could be an important factor in any application which requires the fluid/porous layer to not convect. 

Increasing $\phi$ for fixed $\epsilon$ decreases both $Ra_m$ and $a_m$ as might be expected since a larger macro porosity should mean convective motion is
easier to commence. However, care must be taken in interpretation. The Rayleigh number $Ra_m$ as defined in \eqref{E:RamDef} may be rewritten as
\begin{equation}\label{E:RamDef2}
Ra_m=\frac{\gamma g(T_L-T_U)d_md^2}{\kappa_m\nu}\,\frac{Da}{(1+{\hat d}/{\hat k})}\,\frac{(\rho c)_f}{(\rho c)_m}\,.
\end{equation}
Clearly $Ra_m$ has the correct structure but since $Da=K_f/d^2$ there is a direct dependence on the macro permeability $K^f$ and it is widely believed that
$K^f$ depends strongly on $\phi$. For example, for glass spheres \citet{Chen:1991} uses the Carmen - Kozeny relation
\begin{equation*}
K^f=\frac{d_s^2}{172.8}\,\frac{\phi^3}{(1-\phi)^2}
\end{equation*}
where $d_s$ is the diameter of the glass beads. It is clear that $Ra_m$ in \eqref{E:RamDef2} would then depend strongly on $\phi$ and in any application this 
would need to be taken into account.

We have calculated the necessary values of $a_m$ even though they are not displayed in table \ref{Ta:Ta4}. For the first set of values where
$\phi=0.5$ and $\epsilon=0.2,\cdots,0.6$,
convection always initiates in the porous zone, and $a_m$
takes values 1.4, 1.8, 2.0, 2.1 and 2.2, respectively, while 
$a_m^{(2)}$
has values 18.6 and 18.7 four times. This means lowering the micro porosity increases the width of the convection cells. It is useful to compare this
with the case of a single porosity material as studied in \citet{Straughan:2002ijnamg}. Figure 7 and table VII of that work are appropriate as $\phi=0.5$
there. However, it is difficult to compare directly since the meaning of terms is different, for example, here 
$\delta=\sqrt{K_p}/d_m$
whereas in \citet{Straughan:2002ijnamg},
$\delta=\sqrt{K}/d_m$ with $K$
being the permeability in the single porosity medium. Nevertheless, the case in \citet{Straughan:2002ijnamg} demonstrates convection initiates 
strongly in the porous region. The single porosity case is formally achieved for equations \eqref{E:FluidPert} and \eqref{E:PorPert} by taking
$\xi=0$ and $K_r=1$,
although this is a formal calculation since $\epsilon\to 0$
needs a careful treatment of interface conditions. Although if we take $\xi=0$
and $K_r=1$
and compute $a_m,Ra_m,a_m^{(2)}$ and $Ra_m^{(2)}$
we find for ${\hat d}=0.08$,
$a_m=1.0,Ra_m=3.183,a_m^{(2)}=27.6,Ra_m^{(2)}=5.124$
whereas when ${\hat d}=0.09$,
$a_m=0.7,Ra_m=2.695,a_m^{(2)}=23.6,Ra_m^{(2)}=3.080$.
If these values are compared with corresponding values in table \ref{Ta:Ta1}, for $K_r=1.5$
and table \ref{Ta:Ta4}
then we see that as $\epsilon$
decreases $Ra_m$ decreases and the layer system becomes less stable. This is important because for some applications one requires the layer to
not convect and so the presence of a bidisperse porous medium is highly beneficial. For example, in experiments with a solar pond \citet{Wang:2015}
found that the presence of a porous layer stabilizes the pond and increases efficiency, but, additionally the presence of a bidisperse porous material
further stabilizes and increases efficiency. Of course, our model does not describe a solar pond where one must also account for a salt distribution
and different boundary conditions due to solar radiation heating. However, we do believe it is very interesting that our model does predict
the presence of a bidisperse porous layer could be very beneficial in applications. 

Finally tables \ref{Ta:Ta5} and \ref{Ta:Ta6} show the variation in the values of $Ra_m$ depending on whether the Beavers - Joseph and continuity of normal
stress conditions employ the equally split (ES) value of 0.5 or the pore weighted (PW) values. For the $K_r$ and $\alpha$ values selected here the 
variation is relatively small, but not totally insignificant. It would be useful to have experimental results to compare with in order to assess which
class of interface conditions is preferable.

\section{Conclusions} 

We have formulated equations for thermal convection in a fluid layer which overlies a layer of bidisperse (or double porosity) porous medium
saturated by the same fluid. The macro and micro porosities $\phi$ and $\epsilon$ and the analogous macro and micro permeabilities $K^f$ and $K^p$ are explicitly 
included and play a major role in the model along with the depth of the fluid layer $d$ and the depth of the porous layer $d_m$. The interface conditions
between the fluid and porous medium are very important and we have adopted two approaches. One is to adopt an equal weighting to both the macro and micro phases
while the other approach employs a pore weighted average reflecting the relative porosities $\phi$ and $\epsilon$. 

We suggest a variant of the \citet{BeaversJoseph:1967} interface condition which holds when the porous medium is one of bidispersive type. In general, 
this allows for two Beavers - Joseph interface coefficients $\alpha_1$ and $\alpha_2$ corresponding to the macro and micro phases. In our numerical
results we assume $\alpha_1=\alpha_2$ although in specific applications $\alpha_1$ may need to be different from $\alpha_2$ and our model allows for this.

Our numerical results show that the bidisperse porous medium is very different from the single porous medium case in that the coefficients of the double
porosity material do have a strong effect on convective instability. Since the two layer convection problem with a fluid overlying a bidisperse porous
medium does have serious application to renewable energy generation, see e.g. \citet{Wang:2015,Wang:2018}, we believe the current work is very useful. Of course, two layer
convection with a single porosity medium already has many parameters, see e.g. \citet{Chen:1991}, \citet{ChenChen:1988,ChenChen:1989,ChenChen:1992}, 
\citet{Straughan:2001jcp,Straughan:2002ijnamg}, while convection in a single 
bidisperse layer likewise involves many parameters. The combined problem is necessarily complicated and, therefore, involves a lot of parameters. Future work will
apply this theory to specific renewable energy situations.

\vskip12pt

\noindent{\bf Conflict of interest statement}. This work does not have any conflicts of interest.
 
\vskip12pt

\noindent{\bf Funding}. The work of BS was supported by an Emeritus Fellowship of the Leverhulme Trust, EM-2019-022/9. PD acknowledges partial support by the DFG through project 441523275/SPP225.

\bibliographystyle{plainnat}
\bibliography{thompson}

\begin{thebibliography}{60}
\providecommand{\natexlab}[1]{#1}
\providecommand{\url}[1]{\texttt{#1}}
\expandafter\ifx\csname urlstyle\endcsname\relax
  \providecommand{\doi}[1]{doi: #1}\else
  \providecommand{\doi}{doi: \begingroup \urlstyle{rm}\Url}\fi

\bibitem[Badday and Harfash(2021)]{BaddayHarfash:2021}
A.~J. Badday and A.~J. Harfash.
\newblock Chemical reaction effect on convection in bidispersive porous medium.
\newblock \emph{Transport in Porous Media}, 137:\penalty0 381--397, 2021.

\bibitem[Beavers and Joseph(1967)]{BeaversJoseph:1967}
G.~S. Beavers and D.~D. Joseph.
\newblock Boundary conditions at a naturally permeable wall.
\newblock \emph{J. Fluid Mech.}, 30:\penalty0 197--207, 1967.

\bibitem[Capone and De~Luca(2020)]{CaponeDeLuca:2020}
F.~Capone and R.~De~Luca.
\newblock The effect of the {V}adasz number on the onset of thermal convection
  in rotating bidispersive porous media.
\newblock \emph{Fluids}, 5:\penalty0 5040173, 2020.

\bibitem[Capone and Massa(2021)]{CaponeMassa:2021}
F.~Capone and G.~Massa.
\newblock The effects of {V}adasz term, anisotropy and rotation on bidispersive
  convection.
\newblock \emph{Int. J. Nonlinear Mech.}, 135:\penalty0 103749, 2021.

\bibitem[Capone et~al.(2020{\natexlab{a}})Capone, De~Luca, and
  Gentile]{Capone:2020mrc}
F.~Capone, R.~De~Luca, and M.~Gentile.
\newblock Thermal convection in rotating anisotropic porous layers.
\newblock \emph{Mechanics Res. Comm.}, 110:\penalty0 103601,
  2020{\natexlab{a}}.

\bibitem[Capone et~al.(2020{\natexlab{b}})Capone, De~Luca, and
  Gentile]{Capone:2020prsa}
F.~Capone, R.~De~Luca, and M.~Gentile.
\newblock Coriolis effect on thermal convection in a rotating bidispersive
  porous layer.
\newblock \emph{Proc. Roy. Soc. London A}, 476:\penalty0 20190875,
  2020{\natexlab{b}}.

\bibitem[Capone et~al.(2021)Capone, De~Luca, and Massa]{Capone:2021am}
F.~Capone, R.~De~Luca, and G.~Massa.
\newblock Effect of anisotropy on the onset of convection in rotating
  bidispersive {B}rinkman porous media.
\newblock \emph{Acta Mech.}, 232:\penalty0 3393--3406, 2021.

\bibitem[Carr and Straughan(2003)]{CarrStraughan:2003}
M.~Carr and B.~Straughan.
\newblock Penetrative convection in a fluid overlying a porous layer.
\newblock \emph{Advances in Water Resources}, 26:\penalty0 263--276, 2003.

\bibitem[Chaloob et~al.(2021)Chaloob, Harfash, and Harfash]{Chaloob:2021}
H.~A. Chaloob, A.~J. Harfash, and A.~J. Harfash.
\newblock Bidispersive thermal convection with relatively large macropores and
  generalized velocity and temperature boundary conditions.
\newblock \emph{Phys. Fluids}, 33:\penalty0 014105, 2021.

\bibitem[Chandrasekhar(1981)]{Chandrasekhar:1981}
S.~Chandrasekhar.
\newblock \emph{Hydrodynamic and hydromagnetic stability}.
\newblock Dover, New York, 1981.

\bibitem[Chang et~al.(2006)Chang, Chen, and Straughan]{Chang:2006}
M.~H. Chang, F.~Chen, and B.~Straughan.
\newblock Instability of {P}oiseuille flow in a fluid layer overlying a porous
  layer.
\newblock \emph{J. Fluid Mech.}, 564:\penalty0 287--303, 2006.

\bibitem[Chang et~al.(2017)Chang, Chen, and Chang]{Chang:2017}
T.~Y. Chang, F.~Chen, and M.~H. Chang.
\newblock Stability of plane {P}oiseuille - {C}ouette flow in a fluid layer
  overlying a porous layer.
\newblock \emph{J. Fluid Mech.}, 826:\penalty0 376--395, 2017.

\bibitem[Chen(1991)]{Chen:1991}
F.~Chen.
\newblock Throughflow effects on convective instability in superposed fluid and
  porous layers.
\newblock \emph{J. Fluid Mech.}, 231:\penalty0 113--133, 1991.

\bibitem[Chen and Chen(1988)]{ChenChen:1988}
F.~Chen and C.~F. Chen.
\newblock Onset of finger convection in a horizontal porous layer underlying a
  fluid layer.
\newblock \emph{J. Heat Transfer}, 3:\penalty0 403--409, 1988.

\bibitem[Chen and Chen(1989)]{ChenChen:1989}
F.~Chen and C.~F. Chen.
\newblock Experimental investigation of convective stability in a superposed
  fluid and porous layer when heated from below.
\newblock \emph{J. Fluid Mech.}, 207:\penalty0 311--321, 1989.

\bibitem[Chen and Chen(1992)]{ChenChen:1992}
F.~Chen and C.~F. Chen.
\newblock Convection in superposed fluid and porous layers.
\newblock \emph{J. Fluid Mech.}, 234:\penalty0 97--119, 1992.

\bibitem[Dongarra et~al.(1996)Dongarra, Straughan, and Walker]{Dongarra:1996}
J.~J. Dongarra, B.~Straughan, and D.~W. Walker.
\newblock Chebyshev tau - {QZ} algorithm methods for calculating spectra of
  hydrodynamic stability problems.
\newblock \emph{Appl. Numer. Math.}, 22:\penalty0 399--435, 1996.

\bibitem[Falsaperla et~al.(2016)Falsaperla, Mulone, and
  Straughan]{Falsaperla:2016}
P.~Falsaperla, G.~Mulone, and B.~Straughan.
\newblock Bidispersive inclined convection.
\newblock \emph{Proc. Roy. Soc. London A}, 472:\penalty0 20160480, 2016.

\bibitem[Franchi et~al.(2017)Franchi, Nibbi, and Straughan]{Franchi:2017}
F.~Franchi, R.~Nibbi, and B.~Straughan.
\newblock Continuous dependence on modelling for temperature dependent
  bidispersive flow.
\newblock \emph{Proc. Roy. Soc. London A}, 473:\penalty0 20170485, 2017.

\bibitem[Gentile and Straughan(2017{\natexlab{a}})]{GentileStraughan:2017}
M.~Gentile and B.~Straughan.
\newblock Bidispersive thermal convection.
\newblock \emph{Int. J. Heat Mass Transfer}, 114:\penalty0 837--840,
  2017{\natexlab{a}}.

\bibitem[Gentile and Straughan(2017{\natexlab{b}})]{GentileStraughan:2017prsa}
M.~Gentile and B.~Straughan.
\newblock Bidispersive vertical convection.
\newblock \emph{Proc. Roy. Soc. A}, 473:\penalty0 20170481, 2017{\natexlab{b}}.

\bibitem[Gentile and Straughan(2020)]{GentileStraughan:2020}
M.~Gentile and B.~Straughan.
\newblock Bidispersive thermal convection with relatively large macropores.
\newblock \emph{J. Fluid Mech.}, 898:\penalty0 A14, 2020.

\bibitem[Han et~al.(2020)Han, Wang, and Wang]{Han:2020}
D.~Z. Han, Q.~Wang, and X.~M. Wang.
\newblock Dynamic transitions and bifurcations for thermal convection in the
  superposed free flow and porous media.
\newblock \emph{Physica D}, 414:\penalty0 132687, 2020.

\bibitem[Hibi(2020)]{Hibi:2020}
Y.~Hibi.
\newblock Modelling variable density flow in subsurface and surface water in
  the vicinity of the boundary between a surface water - atmosphere system and
  the subsurface.
\newblock \emph{J. Contaminant Hydrology}, 234:\penalty0 103688, 2020.

\bibitem[Hibi and Tomigashi(2015)]{HibiTomigashi:2015}
Y.~Hibi and A.~Tomigashi.
\newblock Evaluation of a coupled model for numerical simulation of a
  multiphase flow system in a porous medium and a surface fluid.
\newblock \emph{J. Contaminant Hydrology}, 180:\penalty0 34--55, 2015.

\bibitem[Hill and Straughan(2008)]{HillStraughan:2008}
A.~A. Hill and B.~Straughan.
\newblock {P}oiseuille flow of a fluid layer overlying a porous layer.
\newblock \emph{J. Fluid Mech.}, 603:\penalty0 137--149, 2008.

\bibitem[Hill and Straughan(2009{\natexlab{a}})]{HillStraughan:2009}
A.~A. Hill and B.~Straughan.
\newblock Global stability for thermal convection in a fluid overlying a highly
  porous material.
\newblock \emph{Proc. Roy. Soc. London A}, 465:\penalty0 207--217,
  2009{\natexlab{a}}.

\bibitem[Hill and Straughan(2009{\natexlab{b}})]{HillStraughan:2009awr}
A.~A. Hill and B.~Straughan.
\newblock Poiseuille flow in a fluid overlying a highly porous material.
\newblock \emph{Adv. Water Resources}, 32:\penalty0 1609--1614,
  2009{\natexlab{b}}.

\bibitem[Hooman et~al.(2015)Hooman, Sauret, and Dahari]{Hooman:2015}
K.~Hooman, E.~Sauret, and M.~Dahari.
\newblock Theoretical modelling of momentum transfer function of bi-disperse
  porous media.
\newblock \emph{Appl. Thermal Engng.}, 75:\penalty0 867--870, 2015.

\bibitem[Imani and Hooman(2017)]{ImaniHooman:2017}
G.~Imani and K.~Hooman.
\newblock Lattice boltzmann pore scale simulation of natural convection in a
  differentially heated enclosure filled with a detached or attached bidisperse
  porous medium.
\newblock \emph{Trans. Porous Media}, 116:\penalty0 91--113, 2017.

\bibitem[Li et~al.(2018)Li, Xiao, and Lin]{Li:2018}
Y.~Li, S.~Xiao, and Y.~Lin.
\newblock Continuous dependence for the {B}rinkman - {F}orchheimer fluid
  interacting with a {D}arcy fluid in a bounded domain.
\newblock \emph{Math. Comp. Simulation}, 150:\penalty0 66--82, 2018.

\bibitem[Li et~al.(2021{\natexlab{a}})Li, Chen, and Shi]{Li:2021amo}
Y.~Li, X.~Chen, and J.~Shi.
\newblock Structural stability in resonant penetrative convection in a
  {B}rinkman - {F}orchheimer fluid interfacing with a {D}arcy fluid.
\newblock \emph{Appl. Math. Optimization}, 85:\penalty0
  https://doi.org/10.1007/s00245--021--09791--7, 2021{\natexlab{a}}.

\bibitem[Li et~al.(2021{\natexlab{b}})Li, Zhang, and Lin]{Li:2021}
Y.~Li, S.~Zhang, and C.~Lin.
\newblock Structural stability for the {B}rinkman equations interfacing with
  {D}arcy equations in a bounded domain.
\newblock \emph{Boundary Value Problems}, 27:\penalty0
  https://doi.org/10.1186/s13661--021--01501--0, 2021{\natexlab{b}}.

\bibitem[Lyu and Wang(2021)]{LyuWang:2021}
W.~Q. Lyu and X.~M. Wang.
\newblock Stokes - {D}arcy system, small {D}arcy number behaviour and related
  interfacial conditions.
\newblock \emph{J. Fluid Mech.}, 922:\penalty0 A4, 2021.

\bibitem[McCurdy et~al.(2019)McCurdy, Moore, and Wang]{McCurdy:2019}
M.~McCurdy, M.~N. Moore, and X.~Wang.
\newblock Convection in a coupled free flow - porous media system.
\newblock \emph{SIAM J. Appl. Math.}, 79:\penalty0 2313--2338, 2019.

\bibitem[McKay and Straughan(1993)]{McKayStraughan:1993}
G.~McKay and B.~Straughan.
\newblock Patterned ground formation under water.
\newblock \emph{Continuum Mech. Thermodyn.}, 5:\penalty0 145--162, 1993.

\bibitem[Moler and Stewart(1971)]{MolerStewart:1971}
C.~B. Moler and G.~W. Stewart.
\newblock An algorithm for the generalized matrix eigenvalue problem
  ${A}x=\lambda {B}x$.
\newblock Technical report, Univ. Texas at Austin, 1971.

\bibitem[Nield(1977{\natexlab{a}})]{Nield:1977}
D.~A. Nield.
\newblock Onset of convection in a fluid layer overlying a layer of porous
  medium.
\newblock \emph{J. Fluid Mech}, 81:\penalty0 513--522, 1977{\natexlab{a}}.

\bibitem[Nield(1977{\natexlab{b}})]{Nield:2000}
D.~A. Nield.
\newblock Modelling fluid flow and heat transfer in a saturated porous medium.
\newblock \emph{J. Appl. Math. Decis. Sci.}, 81:\penalty0 165--173,
  1977{\natexlab{b}}.

\bibitem[Nield and Kuznetsov(2006)]{NieldKuznetsov:2006}
D.~A. Nield and A.~V. Kuznetsov.
\newblock The onset of convection in a bidisperse of porous medium.
\newblock \emph{Int. J. Heat Mass Transfer}, 49:\penalty0 3068--3074, 2006.

\bibitem[Payne and Straughan(1998)]{PayneStraughan:1998}
L.~E. Payne and B.~Straughan.
\newblock Analysis of the boundary condition at the interface between a viscous
  fluid and a porous medium and related modelling questions.
\newblock \emph{J. Math. Pures Appl.}, 77:\penalty0 317--354, 1998.

\bibitem[Ponalagusamy and Manchi(2021)]{PonalagusamyManchi:2021}
R.~Ponalagusamy and R.~Manchi.
\newblock Mathematical study on two - fluid model for flow of {K}-{L} fluid in
  a stenosed artery with porous wall.
\newblock \emph{SN Applied Sciences}, 3:\penalty0 508, 2021.

\bibitem[Rees(2009)]{Rees:2009}
D.~A.~S. Rees.
\newblock Microscopic modelling of the two - temperature model for conduction
  in heterogeneous media: three - dimensional media.
\newblock In \emph{Proceedings of the fourth International Conference on
  Applications of Porous Media}, volume~13, pages 125--143, Istanbul, 2009.
\newblock ICAPM.

\bibitem[Rees(2010)]{Rees:2010}
D.~A.~S. Rees.
\newblock Microscopic modelling of the two - temperature model for conduction
  in heterogeneous media.
\newblock \emph{J. Porous Media}, 13:\penalty0 125--143, 2010.

\bibitem[Samanta(2020)]{Samanta:2020}
A.~Samanta.
\newblock Linear stability of a plane {C}ouette - {P}oiseuille flow overlying a
  porous layer.
\newblock \emph{Int. J. Multiphase Flow}, 123:\penalty0 103160, 2020.

\bibitem[Saravanan and Vigneshwaran(2020)]{SaravananVigneshwaran:2020}
S.~Saravanan and S.~Vigneshwaran.
\newblock Centrifugal filtration convection in bidisperse media.
\newblock \emph{Phys. Fluids}, 32:\penalty0 084109, 2020.

\bibitem[Sharma and Yadav(2017)]{SharmaYadav:2017}
B.~D. Sharma and P.~K. Yadav.
\newblock A two - layer mathematical model of blood flow in porous constricted
  blood vessels.
\newblock \emph{Transport in Porous Media}, 120:\penalty0 239--254, 2017.

\bibitem[Straughan(2001)]{Straughan:2001jcp}
B.~Straughan.
\newblock Surface - tension - driven convection in a fluid overlying a porous
  layer.
\newblock \emph{J. Computational Phys.}, 170:\penalty0 320--337, 2001.

\bibitem[Straughan(2002)]{Straughan:2002ijnamg}
B.~Straughan.
\newblock Effect of property variation and modelling on convection in a fluid
  overlying a porous layer.
\newblock \emph{Int. J. Numer. Anal. Meth. Geomech.}, 26:\penalty0 75--97,
  2002.

\bibitem[Straughan(2008)]{Straughan:2008}
B.~Straughan.
\newblock \emph{Stability, and wave motion in porous media}, volume 165 of
  \emph{Appl. Math. Sci.}
\newblock Springer, New York, 2008.

\bibitem[Straughan(2017)]{Straughan:2017}
B.~Straughan.
\newblock \emph{Mathematical aspects of multi-porosity continua}, volume~38 of
  \emph{Advances in Mechanics and Mathematics Series}.
\newblock Springer, Cham, Switzerland, 2017.

\bibitem[Straughan(2018)]{Straughan:2018prsa}
B.~Straughan.
\newblock Horizontally isotropic bidispersive thermal convection.
\newblock \emph{Proc. Roy. Soc. London A}, 474:\penalty0 20180018, 2018.

\bibitem[Straughan(2019{\natexlab{a}})]{Straughan:2019abc}
B.~Straughan.
\newblock Anisotropic bidispersive convection.
\newblock \emph{Proc. Roy. Soc. London A}, 475:\penalty0 20190206,
  2019{\natexlab{a}}.

\bibitem[Straughan(2019{\natexlab{b}})]{Straughan:2019prsa}
B.~Straughan.
\newblock Horizontally isotropic double porosity convection.
\newblock \emph{Proc. Roy. Soc. London A}, 475:\penalty0 20180672,
  2019{\natexlab{b}}.

\bibitem[Tiwari et~al.(2020)Tiwari, Shah, and Chauhan]{Tiwari:2020}
A.~Tiwari, P.~D. Shah, and S.~S. Chauhan.
\newblock Solute dispersion in two - fluid flowing through porous tubes with a
  porous layer near the absorbibg wall: {M}odel for dispersion phenomenon in
  microvessels.
\newblock \emph{Int. J. Multiphase Flow}, 131:\penalty0 103380, 2020.

\bibitem[Tsiberkin(2020)]{Tsiberkin:2020}
K.~Tsiberkin.
\newblock Porosity effect on the linear stability of flow overlying a porous
  medium.
\newblock \emph{European Phys. J.}, 43:\penalty0 34, 2020.

\bibitem[Wajihah and Sankar(2021)]{WajihahSankar:2021}
S.~A. Wajihah and D.~S. Sankar.
\newblock Effects of porosity in four - layared nonlinear blood rheology in
  constricted narrow arteries with clinical applications.
\newblock \emph{Computer Methods and Programs in Biomedicine}, 199:\penalty0
  --, 2021.

\bibitem[Wang et~al.(2015)Wang, Yu, Shen, and Zhang]{Wang:2015}
H.~Wang, X.~L. Yu, F.~Shen, and L.~Zhang.
\newblock A laboratory experimental study on effect of porous medium on salt
  diffusion of salt gradient solar pond.
\newblock \emph{Solar Energy}, 122:\penalty0 630--639, 2015.

\bibitem[Wang et~al.(2018)Wang, Zhang, and Mei]{Wang:2018}
H.~Wang, L.~G. Zhang, and Y.~Y. Mei.
\newblock Investigation on the exergy performance of salt gradient solar ponds
  with porous media.
\newblock \emph{Int. J. Exergy}, 25:\penalty0 34--53, 2018.

\bibitem[Yin et~al.(2020)Yin, Wang, and Wang]{Yin:2020}
C.~Yin, C.~W. Wang, and S.~W. Wang.
\newblock Thermal instability of a viscoelastic fluid in a fluid - porous
  system with a plane {P}oiseuille flow.
\newblock \emph{Appl. Math. Mech.}, 41:\penalty0 1631--1650, 2020.

\end{thebibliography}

\vfill\eject

\begin{table}
\begin{center}
\begin{tabular}{|l|l|l|l|l|l|}
\hline
$Kr$   & ${\hat d}$ & $a_m$ & $Ra_m$ & $a_m^{(2)}$ & $Ra_m^{(2)}$ \\ \hline
1.5    & 0.08       & 1.2   & 4.221  & 28.1        & 7.755        \\ 
1.5    & 0.09       & 1.2   & 3.986  & 24.4        & 4.765        \\ 
1.5    & 0.1        & 1.0   & 3.963  & 21.2        & 3.014        \\ \hline
5      & 0.1        & 1.6   & 9.030  & 22.5        & 10.621       \\
5      & 0.11       & 1.6   & 8.746  & 20.1        & 7.172        \\ \hline
10     & 0.11       & 1.8   & 13.901 & 20.4        & 14.604       \\
10     & 0.12       & 1.8   & 13.515 & 18.4        & 10.193       \\ \hline
20     & 0.12       & 2.0   & 19.589 & 18.7        & 20.734       \\ 
20     & 0.13       & 2.0   & 19.128 & 17.0        & 14.852       \\ \hline
25     & 0.12       & 2.1   & 21.474 & 18.7        & 26.015       \\
25     & 0.1260     & 2.1   & 21.178 & 17.7        & 21.276       \\
25     & 0.1261     & 2.1   & 21.173 & 17.7        & 21.206       \\
25     & 0.13       & 2.1   & 20.991 & 17.0        & 18.671       \\ \hline
151.7  & 0.17       & 2.5   & 28.572 & 12.2        & 36.445       \\
151.7  & 0.18       & 2.5   & 28.147 & 11.2        & 27.910       \\ \hline
263.16 & 0.20       & 2.6   & 28.423 & 9.3         & 29.038       \\
263.16 & 0.21       & 2.7   & 27.243 & 8.3         & 22.277       \\ \hline 
\end{tabular}
\caption{
The minimum values of the porous Rayleigh number and corresponding wavenumber for the first minimum, 
$Ra_m,a_m$, and the second minimum, $Ra_m^{(2)},a_m^{(2)}$,
for indicated values of $K_r$. Here, $Pr=6$, $Pr_m=0.75828$, $\delta=0.003279$, ${\hat k}=0.16736$, $\phi=0.3$, $\epsilon=0.3$,
$\alpha=0.1$, $\xi=0.02987$, ${\hat\kappa}=0.12638$. The ${\hat d}$ values are shown in the table.
}%
\label{Ta:Ta1}
\end{center}
\end{table}

\begin{table}
\begin{center}
\begin{tabular}{|l|l|l|l|l|l|}
\hline
$\alpha$ & ${\hat d}$ & $a_m$ & $Ra_m$ & $a_m^{(2)}$ & $Ra_m^{(2)}$ \\ \hline
0.1      & 0.12       & 2.1   & 21.474 & 18.7        & 26.014       \\ 
0.105    & 0.12       & 2.1   & 21.355 & 18.6        & 25.630       \\ 
0.11     & 0.12       & 2.0   & 21.201 & 18.4        & 25.224       \\ 
0.115    & 0.12       & 2.0   & 20.971 & 18.2        & 24.793       \\ 
0.12     & 0.12       & 1.9   & 20.621 & 18.0        & 24.333       \\ 
0.125    & 0.12       & 1.6   & 19.944 & 17.7        & 23.842       \\ 
0.125    & 0.125      & 1.5   & 19.787 & 16.7        & 19.887       \\ 
0.125    & 0.126      & 1.5   & 19.750 & 16.5        & 19.184       \\ \hline 
0.1      & 0.13       & 2.1   & 20.991 & 17.0        & 18.670       \\ 
0.11     & 0.13       & 2.0   & 20.868 & 16.6        & 17.939       \\ 
0.12     & 0.13       & 1.8   & 20.492 & 16.1        & 17.097       \\ \hline 
\end{tabular}
\caption{
The minimum values of the porous Rayleigh number and corresponding wavenumber for the first minimum, 
$Ra_m,a_m$, and the second minimum, $Ra_m^{(2)},a_m^{(2)}$,
for indicated values of the Beavers - Joseph parameter $\alpha$. Here, $Pr=6$, $Pr_m=0.75828$, $\delta=0.003279$, ${\hat k}=0.16736$, 
$\phi=0.3$, $\epsilon=0.3$,
$\xi=0.02987$, ${\hat\kappa}=0.12638$, $K_r=25$. The ${\hat d}$ values are shown in the table.
}%
\label{Ta:Ta2}
\end{center}
\end{table}

\begin{table}
\begin{center}
\begin{tabular}{|l|l|l|l|l|l|}
\hline
$\xi$ & ${\hat d}$ & $a_m$ & $Ra_m$ & $a_m^{(2)}$ & $Ra_m^{(2)}$ \\ \hline
0.1316   & 0.12       & 2.1   & 22.590 & 18.7        & 26.023       \\ 
0.1316   & 0.1245     & 2.1   & 22.345 & 17.9        & 22.367       \\ 
0.1316   & 0.1246     & 2.1   & 22.339 & 17.9        & 22.293       \\ 
0.1316   & 0.125      & 2.1   & 22.318 & 17.9        & 21.999       \\ 
0.1316   & 0.13       & 2.1   & 22.062 & 17.0        & 18.676       \\ \hline 
0.02987  & 0.12       & 2.1   & 21.474 & 18.7        & 26.015       \\ 
0.02987  & 0.125      & 2.1   & 21.226 & 17.9        & 21.992       \\ 
0.02987  & 0.1261     & 2.1   & 21.173 & 17.7        & 21.206       \\ 
0.02987  & 0.1262     & 2.1   & 21.168 & 17.7        & 21.137       \\ 
0.02987  & 0.13       & 2.1   & 20.991 & 17.0        & 18.671       \\ \hline 
\end{tabular}
\caption{
The minimum values of the porous Rayleigh number and corresponding wavenumber for the first minimum, 
$Ra_m,a_m$, and the second minimum, $Ra_m^{(2)},a_m^{(2)}$,
for indicated values of the non-dimensional momentum transfer coefficient $\xi$. Here, $Pr=6$, $Pr_m=0.75828$, $\delta=0.003279$, ${\hat k}=0.16736$, 
$\phi=0.3$, $\epsilon=0.3$,
${\hat\kappa}=0.12638$, $K_r=25$, $\alpha=0.1$. The ${\hat d}$ values are shown in the table.
}%
\label{Ta:Ta3}
\end{center}
\end{table}

\begin{table}
\begin{center}
\begin{tabular}{|l|l|l|l|l|l|l|}
\hline
$\phi$   & $\epsilon$ &${\hat d}$ & $a_m$ & $Ra_m$ & $a_m^{(2)}$ & $Ra_m^{(2)}$ \\ \hline
0.5      & 0.2        & 0.12      & 1.4   & 12.310 & 18.6        & 24.010       \\ 
0.5      & 0.3        & 0.12      & 1.8   & 17.210 & 18.7        & 25.239       \\ 
0.5      & 0.4        & 0.12      & 2.0   & 19.782 & 18.7        & 25.994       \\ 
0.5      & 0.5        & 0.12      & 2.1   & 21.394 & 18.7        & 26.503       \\ 
0.5      & 0.6        & 0.12      & 2.2   & 22.495 & 18.7        & 26.871       \\ \hline 
0.2      & 0.5        & 0.12      & 2.4   & 25.195 & 18.7        & 27.311       \\ 
0.3      & 0.5        & 0.12      & 2.3   & 24.232 & 18.7        & 27.104       \\ 
0.4      & 0.5        & 0.12      & 2.2   & 23.008 & 18.7        & 26.843       \\ 
0.6      & 0.5        & 0.12      & 2.0   & 19.197 & 18.7        & 26.043       \\ \hline 
0.3      & 0.3        & 0.12      & 2.1   & 21.474 & 18.7        & 26.015       \\ 
0.3      & 0.3        & 0.1260    & 2.1   & 21.178 & 17.7        & 21.276       \\ 
0.3      & 0.3        & 0.1261    & 2.1   & 21.173 & 17.7        & 21.206       \\ 
0.3      & 0.3        & 0.13      & 2.1   & 20.991 & 17.0        & 18.671       \\ \hline 
0.5      & 0.2        & 0.14      & 1.3   & 11.692 & 15.6        & 12.909       \\ 
0.5      & 0.2        & 0.1436    & 1.3   & 11.595 & 15.1        & 11.604       \\ 
0.5      & 0.2        & 0.1437    & 1.3   & 11.593 & 15.1        & 11.570       \\ 
0.5      & 0.2        & 0.15      & 1.3   & 11.439 & 14.2        & 9.630        \\ \hline 
0.3      & 0.5        & 0.12      & 2.3   & 24.232 & 18.7        & 27.104       \\ 
0.3      & 0.5        & 0.1234    & 2.3   & 24.054 & 18.1        & 24.097       \\ 
0.3      & 0.5        & 0.1235    & 2.3   & 24.049 & 18.1        & 24.014       \\ 
0.3      & 0.5        & 0.13      & 2.3   & 23.703 & 17.0        & 19.296       \\ \hline 
\end{tabular}
\caption{
The minimum values of the porous Rayleigh number and corresponding wavenumber for the first minimum, 
$Ra_m,a_m$, and the second minimum, $Ra_m^{(2)},a_m^{(2)}$,
for indicated values of the macro porosity $\phi$ and micro porosity $\epsilon$. $Pr=6$, $Pr_m=0.75828$, $\delta=0.003279$, ${\hat k}=0.16736$, 
$\xi=0.02987$,
${\hat\kappa}=0.12638$, $K_r=25$, $\alpha=0.1$. The ${\hat d}$ values are shown in the table.
}%
\label{Ta:Ta4}
\end{center}
\end{table}

\begin{table}
\begin{center}
\begin{tabular}{|l|l|l|l|l|l|}
\hline
$Kr$   & ${\hat d}$ & $Ra_m$(ES) & $Ra_m$(PW) & $Ra_m^{(2)}$(ES) & $Ra_m^{(2)}$(PW) \\ \hline
20     & 0.12       & 19.589     & 18.199     & 20.734           & 20.997           \\ 
20     & 0.13       & 19.128     & 17.666     & 14.852           & 15.216           \\ \hline
25     & 0.12       & 21.474     & 20.132     & 26.015           & 26.310           \\ 
25     & 0.13       & 20.991     & 19.539     & 18.671           & 19.103           \\ \hline 
151.7  & 0.17       & 28.572     & 28.048     & 36.445           & 39.233           \\ 
151.7  & 0.18       & 28.147     & 27.632     & 27.910           & 30.664           \\ 
151.7  & 0.19       &            & 27.113     &                  & 24.152           \\ \hline 
263.16 & 0.19       & 29.013     &            & 37.655           &                  \\ 
263.16 & 0.20       & 28.423     & 28.340     & 29.038           & 33.579           \\ 
263.16 & 0.21       & 27.243     & 27.790     & 22.277           & 26.821           \\ \hline 
\end{tabular}
\caption{
The minimum values of the porous Rayleigh number and corresponding wavenumber for the first minimum, 
$Ra_m$, and the second minimum, $Ra_m^{(2)}$,
for a comparison between the equally split (ES) and pore weighted (PW) versions of the Beavers - Joseph interface condition, and the continuity of normal 
stress at the interface. Here $Pr=6$, $Pr_m=0.75828$, $\delta=0.003279$, ${\hat k}=0.16736$, 
$\phi=0.3$, $\epsilon=0.3$,
${\hat\kappa}=0.12638$, $\xi=0.02987$, $\alpha=0.1$. The ${\hat d}$,  $K_r$ values are shown in the table.
}%
\label{Ta:Ta5}
\end{center}
\end{table}

\begin{table}
\begin{center}
\begin{tabular}{|l|l|l|l|l|l|}
\hline
$\alpha$   & ${\hat d}$ & $Ra_m$(ES) & $Ra_m$(PW) & $Ra_m^{(2)}$(ES) & $Ra_m^{(2)}$(PW) \\ \hline
0.1        & 0.12       & 21.474     & 20.132     & 26.014           & 26.310           \\ 
0.105      & 0.12       & 21.355     & 20.018     & 25.630           & 26.034           \\ 
0.11       & 0.12       & 21.201     & 19.867     & 25.224           & 25.745           \\ 
0.115      & 0.12       & 20.971     & 19.681     & 24.793           & 25.442           \\ 
0.12       & 0.12       & 20.621     & 19.442     & 24.333           & 25.123           \\ 
0.125      & 0.12       & 19.944     & 19.115     & 23.842           & 24.788           \\ 
0.13       & 0.12       &            & 18.608     &                  & 24.434           \\ 
0.135      & 0.12       &            & 17.659     &                  & 24.059           \\ \hline 
0.1        & 0.13       & 20.991     & 19.539     & 18.670           & 19.103           \\ 
0.11       & 0.13       & 20.868     & 19.372     & 17.939           & 18.584           \\ 
0.12       & 0.13       & 20.492     & 19.075     & 17.097           & 18.005           \\ 
0.13       & 0.13       &            & 18.352     &                  & 17.352           \\ \hline 
\end{tabular}
\caption{
The minimum values of the porous Rayleigh number and corresponding wavenumber for the first minimum, 
$Ra_m$, and the second minimum, $Ra_m^{(2)}$,
for a comparison between the equally split (ES) and pore weighted (PW) versions of the Beavers - Joseph interface condition, and the continuity of normal 
stress at the interface. 
Here $Pr=6$, $Pr_m=0.75828$, $\delta=0.003279$, ${\hat k}=0.16736$, 
$\phi=0.3$, $\epsilon=0.3$,
${\hat\kappa}=0.12638$, $\xi=0.02987$, $K_r=25$. The ${\hat d}$,  $\alpha$ values are shown in the table.
}%
\label{Ta:Ta6}
\end{center}
\end{table}

\vfill\eject

\end{document}